\documentclass[%
twocolumn,aps, prb, 
superscriptaddress,
 amsmath,amssymb,
]{revtex4-2}

\usepackage{graphicx}% Include figure files
\usepackage{dcolumn}% Align table columns on decimal point
\usepackage{bm}% bold math
\usepackage{xcolor}
\usepackage{cancel}
\usepackage[colorlinks, linkcolor=red,citecolor=blue,urlcolor=blue]{hyperref}
\usepackage[all]{hypcap} % let hyperlinks correctly point to figures rather than their captions;
\begin{document}

\title{
Strongly nonlinear regime of Josephson transmission lines revealed by two-tone spectroscopy
}

\author{A.~S.~Averkin}
\affiliation{National University of Science and Technology ``MISIS'', 119049 Moscow, Russia}
\author{A.~A.~Kopasov}
\affiliation{National University of Science and Technology ``MISIS'', 119049 Moscow, Russia}
\author{I.~E.~Pologov}
\affiliation{National University of Science and Technology ``MISIS'', 119049 Moscow, Russia}
\author{Aleksey~N.~Bolgar}
\affiliation{Moscow Institute of Physics and Technology, Dolgoprudny 141700, Russia}
\author{Daria~A.~Kalacheva}
\affiliation{Moscow Institute of Physics and Technology, Dolgoprudny 141700, Russia}
\affiliation{Skolkovo Institute of Science and Technology, Skolkovo Innovation Center, Moscow 121205, Russia}
\author{Viktor B. Lubsanov}
\affiliation{Moscow Institute of Physics and Technology, Dolgoprudny 141700, Russia}
\author{M.~V.~Fistul}%
\affiliation{National University of Science and Technology ``MISIS'', 119049 Moscow, Russia}
\author{A.~Karpov}
\affiliation{National University of Science and Technology ``MISIS'', 119049 Moscow, Russia}

\date{\today}% It is always \today

\begin{abstract}

We present experimental and theoretical studies of the off-resonant and strongly nonlinear regime of Josephson transmission lines (JTLs) with galvanically-coupled nonlinear elements. The transition from the weakly to the strongly nonlinear regime of a JTL induced by increasing the input power of the pump is probed via two-tone spectroscopy. Measurements of the phase of the transmission coefficient for a weak probe signal reveal a large increase and pronounced oscillations in the phase length variation as a function of the microwave power of the pump. Experimental observations are explained on the basis of the developed theoretical approach suitable for the description of the nonlinear response of strongly driven JTLs. Using the derived nonlinear wave equation, we show that the behavior of the phase length variation is associated with the oscillatory dependence of the Josephson inductances on the microwave power. It is demonstrated that the dissipation-induced propagation losses increase in the strongly nonlinear regime and also lead to smearing out the phase length oscillations. The developed theoretical analysis is in good agreement with experimental observations.

\end{abstract}

\maketitle

\section{\label{sec:level1}Introduction}

Superconducting metamaterials have attracted considerable attention from both theoretical and experimental groups for several decades~\cite{ricci2005superconducting,anlage2010physics,Jung2014,Lazarides}. Such systems typically consist of networks of lumped superconducting circuits coupled to a low-dissipative transmission line or a resonator~\cite{ricci2005superconducting,Lazarides, anlage2010physics,Jung2014,butz2013one,Multistability,Kiselev,rotzinger2016aluminium,wildermuth2022fluxons,maleeva2018circuit,zapata2024granular}. Exploiting the intrinsic nonlinearity of Josephson junctions or the kinetic inductance of superconducting films allowed observation of a number of phenomena including dynamic metastable states~\cite{Multistability}, magnetic fluxon propagation~\cite{wildermuth2022fluxons}, the current resonances~\cite{barbara1999stimulated}, the giant Shapiro steps~\cite{halsey1990giant}, and the parametric amplification~\cite{Lehnert,Eom2012}. These effects, in turn, form basic operation principles for various devices, e.g., the low-temperature parametric amplifiers~\cite{Eom2012,Lehnert} and the voltage standard~\cite{kohlmann2011development}.  

Systems based on Josephson junction or superconducting interference device (SQUID) arrays are a promising experimental platform to realize a nonlinear optical medium and to establish precise control over the microwave radiation~\cite{Lazarides,Jung2014}. Experimental research in this field encompasses both one-dimensional~\cite{Jung,Butz} and two-dimensional SQUID arrays~\cite{Anlage2013,Averkin2016,Anlage2017,Zhuravel2019}. A prominent class of superconducting metamaterials incorporates the systems with chains of Josephson junctions~\cite{Siddiqi, Brien,Macklin, WhiteAPL2015,malnou2025travelling}, SQUIDs~\cite{Zorin, MianoIEEE2019, Grunhaupt}  or superconducting nonlinear asymmetric elements (SNAILs)~\cite{BellPRAppl2015, RoudsaryAPL2023, RevKerr} \textit{directly embedded} into the transmission line. Hereafter, we refer to these metamaterials as the Josephson transmission lines (JTLs). There is strong interest in the physics of JTLs because this class represents a typical design of the Josephson traveling-wave parametric amplifier. Depending on the structure and flux biasing, the amplification of a weak signal propagating through a JTL in the presence of a relatively strong drive can be achieved via the four- or three-wave mixing processes allowed by the lowest-order cubic (Kerr-type) or quadratic nonlinearity of the Josephson inductance, respectively~\cite{Siddiqi, Zorin}. The study of propagation and mixing of electromagnetic waves in JTLs is also important for several research fields including the generation of nonclassical states of light~\cite{GrimsmoNPJ2017, EspositoPRL2022, PerelshteinPRAppl2022, QiuNP2023}, dark matter searches~\cite{BartramRSI2023,DiVoraPRD2023,RamanathanPRL2023}, and the readout from microwave single-photon detectors~\cite{BockstiegelJLTP2014,ZobristAPL2019}.

Usually, the Josephson traveling-wave parametric amplifier operates in the so-called off-resonant regime, namely, at frequencies $\omega$ that are much smaller than typical resonant frequencies of the unit cell $\omega_{{\rm res}}$ ($\omega \ll \omega_{\rm res}$). Characteristic amplitudes of the pump tones are not large, so that the typical microwave current in the transmission line $I$ is much smaller than the critical current of the Josephson junctions $I\ll I_c$.
Consequently, the vast majority of experimental and theoretical works on Josephson traveling-wave amplifiers address the off-resonant and weakly nonlinear regime, which is described reasonably well by the coupled mode equations approach (see Refs.~\cite{WhiteAPL2015, malnou2025travelling} and references therein). At the same time, the behavior of JTLs in the opposite, \textit{strongly nonlinear}, regime is largely unexplored. Motivation for considering this case stems from a natural expectation that the emerging nonlinear effects can be enhanced upon the increase in the power of the pump tone. In addition, it is largely anticipated that the propagation losses should increase in the strongly nonlinear regime causing gain compression. Nevertheless, to the best of our knowledge, the detailed experimental and theoretical studies addressing the strongly nonlinear regime of JTLs are lacking. The main goal of our work is to fill these gaps and provide both experimental and theoretical studies of the propagation of electromagnetic waves in JTLs within a wide range of pump powers. 

From the experimental viewpoint, the standard technique for addressing the behavior of the SQUID arrays coupled to a transmission line is the measurement of the scattering 
parameters in the single- and two-tone regimes. In particular, such a technique was used in studies of SQUID arrays weakly coupled to a transmission line~\cite{butz2013one,Multistability,Snake,Kiselev,Intermodulation} and allowed direct observations of several phenomena including oscillations of the resonant frequency under the action of a single microwave signal~\cite{Snake}, and the response of two-dimensional Josephson transmission line to two-tone excitation~\cite{Intermodulation}. Measurements of the $S$ parameters are also routinely used for characterization of Josephson traveling-wave amplifiers. In particular, the behavior of the phase of the transmission coefficient, ${\rm arg}\left(S_{21}\right)$, in the single- and two-tone regimes as a function of the pump power provides important information about the self- and cross-phase modulation in such devices (see, e.g., Ref.~\cite{Grunhaupt} for a recent work). Experimentally, the strongly nonlinear regime implies the JTL pump power to be big enough for the departure from the monotonic growth of the phase length versus pump power.

\begin{figure*}[htpb]
\includegraphics[scale=0.32]{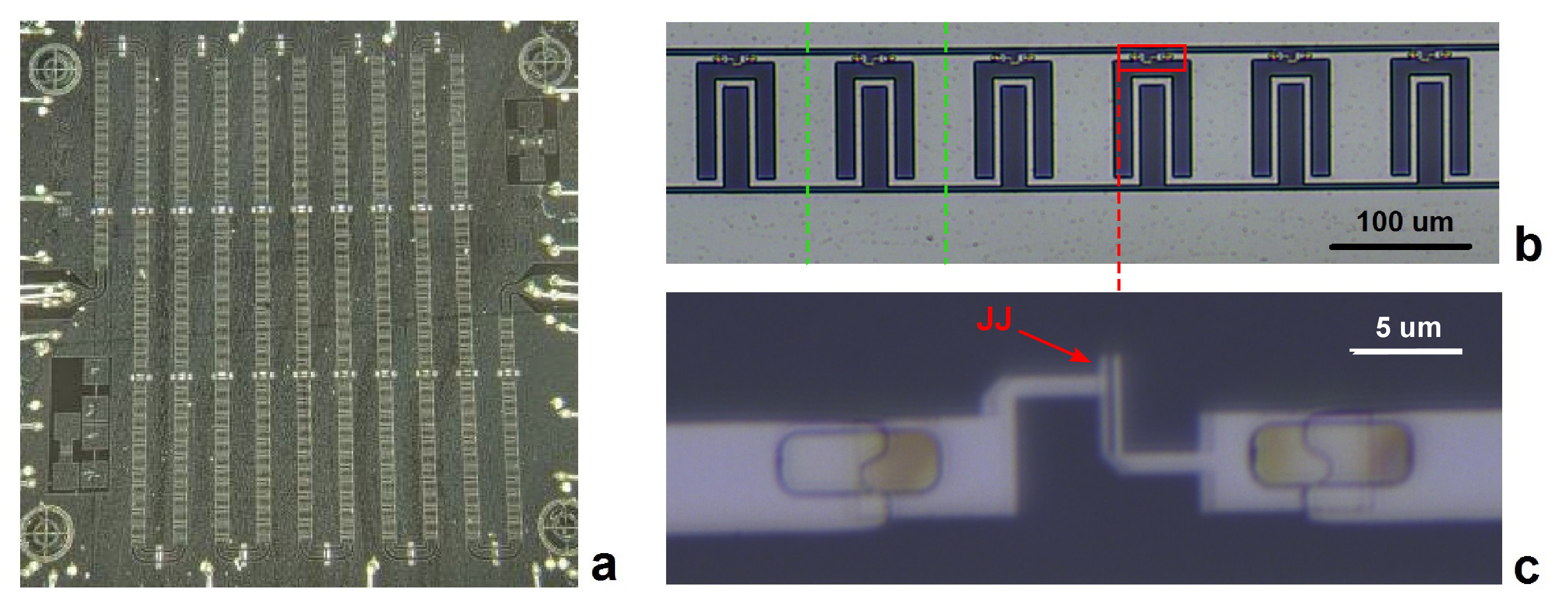}
\caption{\label{fig:Sample} (a) Photo of 5$\times$5~mm chip containing the chain of $N=356$ SQUIDs. Input and output contact pads are placed on the left and right sides of the chip, correspondingly. (b) Photo of the part of the chip showing six SQUIDs directly embedded into the central conductor of coplanar waveguide. Each SQUID consists of the two branches: the first branch has the meander shape, while the second one is interrupted by a Josephson junction (marked by red rectangle). The meander branch of the rf-SQUID corresponds to the geometric inductance $L_g$, while the segment of the coplanar waveguide together with the ground plane provides the capacitance $C_0$. The unit cell of sample (between two green dashed lines) consists of an rf-SQUID and a segment of the coplanar waveguide. (c) Close view of Josephson junction embedded into SQUID's branch fabricated by shadow evaporation. The Josephson junction (JJ) is shown by red arrow. In (b) and (c) the Al conductors appear as bright regions, while the dark regions correspond to the substrate.}
\end{figure*}

Theoretical analysis of the wave propagation and mixing in JTLs usually employs two approaches: numerical simulations~\cite{DIxonPRAppl2020,GaydamachenkoJAP2022,LevochkinaSUST2024} and coupled-mode equations~\cite{Siddiqi,Brien,Zorin,malnou2025travelling,BellPRAppl2015}. Numerical methods are applicable in the strongly nonlinear regime and allow one to calculate the $S$ parameters and identify the dominating processes in the nonlinear response. However, these methods are not particularly informative. Conversely, the coupled-mode equations approach is more transparent but cannot be directly applied for the analysis of the JTL response in the strongly nonlinear regime. So, from the analytical point of view, the strongly nonlinear regime model requires an approximation beyond the quadratic and/or cubic nonlinearity in order to explain the particular details of the phase behavior. An important idea, which can be useful for developing  an appropriate theoretical description for strongly driven JTLs, is the one used in Refs.~\cite{Multistability,Kiselev} for the analysis of strongly driven SQUID chains with rather weak inductive coupling to the transmission line. This idea is based on the observation that in the strongly nonlinear regime the response of the superconducting metamaterial is dominated by the signal at the frequency of the drive. Nevertheless, there is a crucial difference between the systems studied in Refs.~\cite{Multistability,Kiselev} and the ones considered in the present work, which lies in the presence of a strong galvanic coupling between nonlinear elements of a JTL. Moreover, in Refs. ~\cite{Multistability,Kiselev} the crucial assumption that the amplitudes of the microwave signal and pump tones do not vary along the sample was used. In this regard, developing a theoretical framework for strongly driven JTLs is not a straightforward task and requires the generalization of the approach used in Refs.~\cite{Multistability, Kiselev}.

The paper is organized as follows. In Sec.~\ref{subsection:Sec_2_device} we provide a detailed description of the physical device and an experimental setup for the two-tone spectroscopy measurements. In Sec.~\ref{subsection:Sec_2_expresults} experimental results, i.e., measurements of the dependence of the phase length variation for a signal tone on the microwave power of a pump tone are presented and discussed. In Sec.~\ref{subsection:Sec_3_model} we introduce a basic electromagnetic model of an array of rf-SQUIDs directly embedded into the transmission line and derive the nonlinear wave equation, which we use to study the propagation of electromagnetic waves in strongly driven JTLs. In Sec.~\ref{subsection:Sec_3_single} and~\ref{subsection:Sec_3_Two_Tone} we apply our theory to JTLs in the presence of  one and two microwave tones. Our main results are summarized and discussed in Sec.~\ref{section:Sec_6_discussion}. Appendix~\ref{boundary_condition_derivation} contains the derivation of the boundary condition for equations of the electrodynamic model.
Appendix~\ref{modified_equations_two_tones_appendix} contains the details of the derivations of the basic equations for the two-tone spectroscopy. Appendix~\ref{experimental_gain} presents additional experimental results on the behavior of $|S_{21}|$ in the strongly nonlinear regime. Appendix~\ref{app:third_harmonic} contains the analysis of third-harmonic generation and discusses the limitations of the one-harmonic approximation.

\section{\label{sec:Sec_2_experiment} Phase of the transmission coefficient $S_{21}$ in JTL\lowercase{s}: experiment}
\subsection{\label{subsection:Sec_2_device} Physical device and experimental setup}

The key element of the sample is a single-Josephson junction SQUID (rf-SQUID). The rf-SQUID superconducting loop consists of the two superconducting wires: one branch is interrupted by the Josephson junction with the critical current $I_c=1.45$~$\mu$A, while the second branch has the meander shape in order to obtain a large geometric inductance per unit cell $L_g=116$~pH (Fig. \ref{fig:Sample}). The capacitance of the Josephson junction is $C_J = 35$~fF. The Josephson inductance at zero phase difference $L_J = \Phi_0/(2 \pi I_c)= 227$~pH, where $\Phi_0$ is the magnetic flux quantum. The corresponding rf-SQUID screening parameter $\beta_L=L_g/L_J = 0.5$. Each SQUID is directly embedded into the central line of coplanar waveguide (CPW). Capacitance per unit cell between the central conductor of CPW and the ground plane $C_0 = 31$~fF is designed to be large in order to obtain a characteristic impedance of CPW close to $Z_0=50$~Ohm. The unit cell has the length $a=106$~$\mu$m, and the sample consists of $N=356$ unit cells.  The Al coplanar line and the ground plane were fabricated on a silicon chip using the e-beam evaporation, optical lithography and dry etching processes. Al/AlO$_x$/Al Josephson junctions were fabricated using e-beam lithography and the shadow-evaporation technique at the Plasys facility. A photograph of the sample is shown in Fig.~\ref{fig:Sample}. The equivalent circuit is shown in Fig.~\ref{fig:circuit1}.

\begin{figure}[tb]
\centering
\includegraphics[scale=0.24]{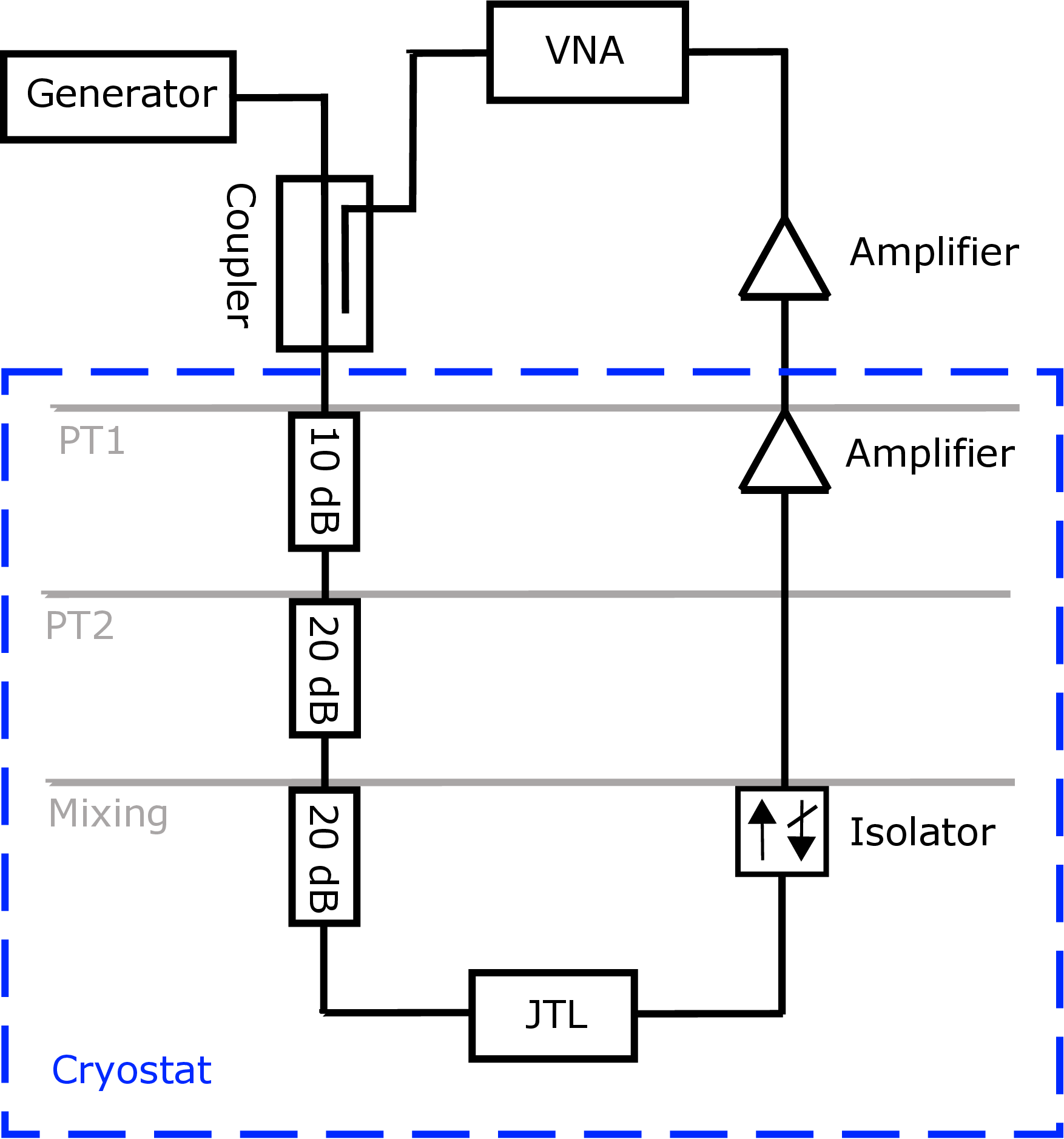}
\caption{\label{fig:Setup} Experimental setup for the two-tone spectroscopy measurements.}
\end{figure}

The experimental setup is shown in Fig.~\ref{fig:Setup}. The sample was mounted inside the sample holder using Al wire-bonds. The sample holder was placed inside the magnetic shield from Cryoperm and installed in an Oxford Instruments DR200 dilution refrigerator. Two coaxial lines were used to apply RF signals to the sample. A set of microwave attenuators providing a total attenuation of 50~dB was mounted inside the cryostat on the input RF line. The output RF line passes through an RF isolator, low noise amplifier (LNF-LNC0.3-14A) and the room temperature amplifier (ZVA-183-S+). Experiment was performed at a base temperature of the cryostat $T \approx 40$~mK. The transmission coefficient $S_{21}$ was measured using an Agilent PNA-X Vector Network Analyzer (VNA). Pump tone was applied from the signal generator (Agilent E8257D) via the RF coupler.

\subsection{\label{subsection:Sec_2_expresults}Two-tone spectroscopy: phase length measurements}

We measured the transmission coefficient $S_{21}$ of a signal tone as a function of the power of a pump tone. To realize this, we employ two-tone spectroscopy, where the sample was irradiated by two microwave tones: a weak signal tone at frequency $f_s$ and a strong pump tone at frequency  $f_p$.
The pump tone was formed by a signal generator at fixed frequency $f_p=6$~GHz, while a Vector Network Analyzer was used to measure the complex transmission coefficient $S_{21}$ versus the signal frequency. The phase difference of $S_{21}$ (\textit{the phase length variation}) for the signal tone at frequency $f_s$ was extracted from the measured transmission coefficient $-{\rm arg}(S_{21})$ normalized to the phase at a low pump power reference level ($-75$~dBm).
The dependence of the phase length variation for a signal tone on the power of the microwave pump tone was measured for three signal frequencies $f_s$: $3$, $6.1$ and $9$~GHz, while the pump tone frequency was maintained constant at $f_p=6$~GHz (Fig.~\ref{fig:TT}(a)). Fig.~\ref{fig:TT}(b) shows typical behavior of $|S_{21}|$ as a function of the pump power. It can be seen that the studied JTL exhibits rather moderate gain within a power range of approximately 10 dBm (from -57 to -47 dBm).

In comparison to $|S_{21}|$, which doesn't reveal any peculiarities, the phase length variation has a number of specific features in the strongly nonlinear regime, the explanation of which represents the goal of this work. All three curves in Fig.~\ref{fig:TT}(a) display a large increase and pronounced oscillations as the power of the pump tone increases. For example, at the signal frequency $f_s=9$~GHz the measured variation of the phase length increases with pump power, reaching the maximum value of approximately $500^\circ$ at the pump power $-48$~dBm. For higher pump powers, the phase length starts to oscillate, showing two distinct decaying oscillations before stabilizing at approximately $400^\circ$. One can see that both the maximum magnitude and the amplitude of the oscillations increase with the signal frequency, $f_s$.

The experimentally observed behavior of the phase-length variation as a function of pump power can be understood qualitatively as follows. With increasing pump amplitude, the phase oscillations across the Josephson junctions become large, which leads to the suppression of the effective Josephson contribution to the propagation of the wave at the pump frequency. As a result, the effective inductance of a unit cell increases from $L_{\rm eff} = L_gL_J/(L_g + L_J)$ toward $L_{\rm eff} = L_g$. Since the propagation velocity scales as $v = \omega/k \propto 1/\sqrt{L_{\rm eff}C_0}$, this increase in the effective inductance leads to a decrease in the wave velocity. At fixed frequency this corresponds to an increase in the local wave number $k$ and, consequently, to a larger phase length. In the strongly nonlinear regime this renormalization of the unit-cell inductance becomes oscillatory as a function of the pump amplitude because the Josephson response is governed by Bessel-function factors. This gives rise to the oscillatory dependence of the phase length on pump power.

\section{\label{section:Sec_3_theory} Phase of the transmission coefficient $S_{21}$ in JTL\lowercase{s}: theory
}
\subsection{\label{subsection:Sec_3_model} Electrodynamic model and nonlinear wave equation}

\begin{figure}[!tb]
\centering
\includegraphics[scale = 0.72]{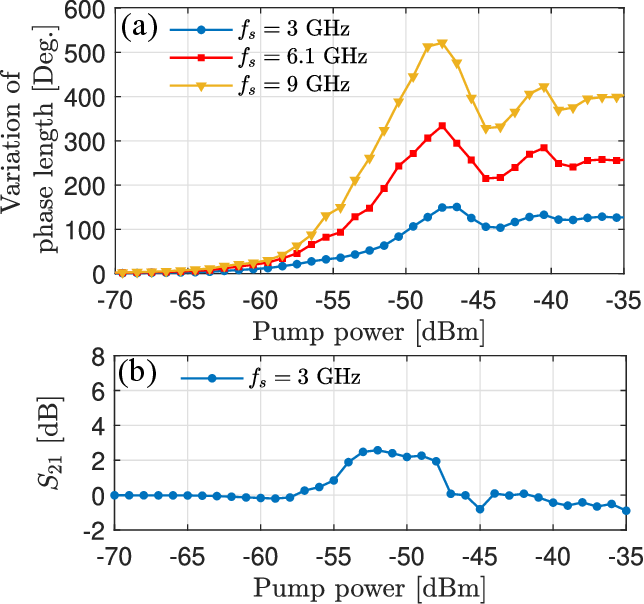}
\caption{\label{fig:TT} Measured two-tone spectroscopy. (a) Typical dependencies of the phase length variation for a weak probe tone at frequencies $f_s= 3$, $6.1$, and $9$~GHz as a function of the input power of the pump at frequency $f_p=6$~GHz. (b) Behavior of the transmission $|S_{21}|$ as a function of the pump power for $f_p = 6$~GHz and $f_s = 3$~GHz.}
\end{figure}

\begin{figure*}[t]
\centering
\includegraphics[scale =0.62]{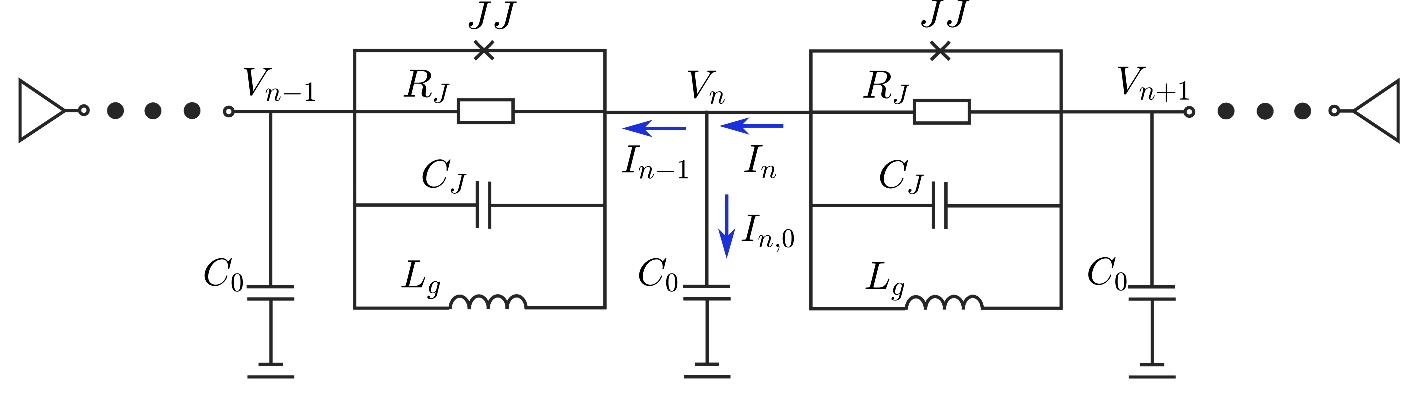}
\caption{\label{fig:circuit1} Josephson transmission line scheme. Josephson junctions, capacitors, resistors, and inductances are denoted by $JJ$, $C$, $R$, and $L$, respectively. The node voltages and currents are denoted by $V_n$ and $I_n$, respectively. The length of the unit cell is equal to $a$.
}
\end{figure*}

To address the basic features of experimental observations reported in the previous section, we proceed with the discussion of the electromagnetic model. The equivalent circuit of the considered JTL is shown in Fig.~\ref{fig:circuit1}. The external magnetic flux is taken to be zero. Each unit cell is characterized by the following parameters: the critical current of the Josephson junction at zero magnetic field $I_c$, the Josephson junction capacitance $C_J$, and the geometrical inductance $L_g$. For simplicity, the transmission line is assumed to be lossless and characterized by capacitance to the ground $C_0$. We also assume that the self-inductance of a transmission line is much less than $L_g$ and $L_J$. The dissipation-induced propagation losses are modelled by the ohmic resistor $R_J$.    

The classical nonlinear dynamics of such JTLs is completely determined by a set of time-dependent node phases (dimensionless node-fluxes \cite{leib2012networks}), $\phi_n(t)$, with $n$ varying from $0$ up to $N=\ell/a$, where $\ell$ is the length of the chain. Using Kirchhoff's circuit laws and the two Josephson relations we write down a set of nonlinear dynamic equations governing the classical electrodynamics of JTLs.  
 \begin{equation}
 I_n-I_{n+1}=C_0\frac{dV_n}{dt}
 \label{CurrentNode}
\end{equation}
and 
 \begin{equation}
 V_n=\frac{\Phi_0}{2\pi}\frac{d\phi_n}{dt}.
\label{f2}
 \end{equation}
The current $I_n$ flowing through the $n$-th cell is written as
 \begin{eqnarray}
 \label{f3}
I_{n}=C_J\frac{\Phi_0}{2\pi}(\ddot\phi_n-\ddot\phi_{n-1})+\frac{\Phi_0}{2\pi R_j}(\dot\phi_n-\dot\phi_{n-1})\\
\nonumber
+\frac{\Phi_0}{2\pi L_g} (\phi_n-\phi_{n-1})+I_c \sin(\phi_{n}-\phi_{n-1}) \ .
\end{eqnarray}
All the relevant physical quantities, i.e., node currents $I_n$ and voltages $V_n$, are indicated in Fig. \ref{fig:circuit1}. Combining Eqs.~(\ref{CurrentNode})-(\ref{f3}), we obtain a set of coupled nonlinear differential equations for dynamic variables $\phi_n(t)$ 
\begin{eqnarray}\label{main_equations_discrete}
C_0\ddot{\phi}_n-C_J\left(\ddot{\phi}_{n+1}+\ddot{\phi}_{n-1}-2\ddot{\phi}_n\right)- ~~~~~~~~~~~~~~~~~~~~~~~~~ \nonumber \\
- \frac{(\phi_{n+1}+\phi_{n-1}-2\phi_n)}{L_g}
 - \frac{1}{L_J}\sin\left(\phi_{n+1}-\phi_n\right) + ~~~~~~~~~\nonumber \\
  + \frac{1}{L_J}\sin\left(\phi_n-\phi_{n-1}\right) 
  - \frac{(\dot{\phi}_{n+1}+\dot{\phi}_{n-1}-2\dot{\phi}_n)}{R_J} = 0. ~~~~~~~ 
\end{eqnarray}
Since for a typical range of microwave frequencies the condition $ka \ll 1$ ($k$ denotes the wavenumber of electromagnetic waves propagating in a JTL) is satisfied, one can pass to the continuum limit in Eq.~(\ref{main_equations_discrete}) 
\begin{subequations}\label{cl_transformation}
\begin{align}
\phi_{n+1} + \phi_{n-1}-2\phi_n \approx  \frac{\partial^2\phi}{\partial x^2} \ ,\\
\sin(\phi_{n+1}-\phi_n)-\sin(\phi_n - \phi_{n-1}) \approx 
\nonumber \\
\approx  2\cos\left(\frac{\partial\phi}{\partial x}\right)\sin\left(\frac{1}{2}\frac{\partial^2\phi}{\partial x^2}\right) \ ,
\end{align}
\end{subequations}
and obtain the\textit{ nonlinear wave equation } for the coordinate- and time-dependent phase, $\phi(x,t)$:
\begin{eqnarray}\label{main_equations_continuous}
C_0\ddot{\phi} - \frac{1}{L_g}\frac{\partial^2\phi}{\partial x^2} - \frac{1}{R_J}\frac{\partial^2 \dot{\phi}}{\partial x^2} = 
\nonumber \\
=\frac{2}{L_J}\cos\left(\frac{\partial\phi}{\partial x}\right) \sin\left(\frac{1}{2}\frac{\partial^2\phi}{\partial x^2}\right) \ .
\end{eqnarray}
Hereafter, the coordinate $x$ is normalized to the length of the unit cell. It is important to note that in
writing the transformations~(\ref{cl_transformation}) and the dynamic equation~(\ref{main_equations_continuous}) we neglected small corrections associated with deviations of the dispersion relation from the linear one, which is justified in the off-resonant case $\omega \ll \omega_0$, $\omega_J$, where $\omega_0 = 1/\sqrt{C_0L}$ and $\omega_J = 1/\sqrt{C_JL}$ is the rf-SQUID plasma frequency, and $L^{-1} = L_J^{-1} + L_g^{-1}$ is the inverse total inductance of the unit cell. Moreover, for the considered frequency and parameter range, the junction capacitance $C_J$ doesn't affect the dynamics and can be disregarded.

Equation~(\ref{main_equations_continuous}) forms the basis for our theoretical analysis of phase length measurements. In the rest of this section we derive and solve approximate equations describing the propagation of electromagnetic waves inside the system in the strongly nonlinear regime. Note that theoretical description of the strongly nonlinear regime realized in such JTLs goes beyond the Kerr-type nonlinearity well known in quantum optics~\cite{ferretti2012single,brauer2021generation,maleeva2018circuit}.
If not stated otherwise, the results of our theoretical approach presented in this section are obtained for the following set of circuit parameters: $L_J = 227$~pH, $L_g = 116$~pH, $C_0 = 31$~fF, and the chain consists of $N = 356$ unit cells. For this parameter set the total inductance $L  = 76.6$~pH, and the inductance ratio $L/L_J =  0.34$. We achieve the best fit to experimental data for $R_J = 180$~Ohm. 

\subsection{\label{subsection:Sec_3_single} Single-tone spectroscopy}

Here, we quantitatively analyze a single-tone spectroscopy of JTLs biased in the off-resonant regime. This analysis will form the basis for the description of the two-tone spectroscopy. In this setup, a JTL is subjected to an externally applied microwave radiation of the frequency $\omega$ and power $P$.  Our starting point is Eq.~(\ref{main_equations_continuous}) and inspired by the approach of Ref.~\cite{Multistability,Kiselev}, we look for the solution in the form $\phi(x,t) = A(x)\cos[\omega t - \theta(x)]$. 
The spatial derivatives of the node phases in the nonlinear wave equation~(\ref{main_equations_continuous}) can be rewritten as 
\begin{subequations}\label{derivatives_expressions}
\begin{align}
\frac{\partial\phi}{\partial x} = \frac{d A}{dx}\cos(\xi) + A\frac{d\theta}{dx}\sin(\xi) \ ,\\
\label{node_flux_second_derivative}
\frac{\partial^2\phi}{\partial x^2} = \left[\frac{d^2A}{dx^2} - A\left(\frac{d\theta}{dx}\right)^2\right]\cos(\xi) 
\nonumber \\
 + \left(A\frac{d^2\theta}{dx^2} + 2\frac{dA}{dx}\frac{d\theta}{dx}\right)\sin(\xi) \ ,
\end{align}
\end{subequations}
where $\xi(x,t) = \omega t - \theta(x)$ and in the following we use a compact notation for spatial derivatives $df/dx = f'$. The $\theta'$ can be interpreted as the local wave number and in the considered low-frequency range $\theta' \ll 1$. In the presence of a weak dissipation 
one can neglect $A''$ in comparison to $A(\theta')^2$ and $A\theta''$ in comparison to $A'\theta'$ in the above expressions. Using Eqs.~(\ref{derivatives_expressions}) and linearizing the right-hand-side of Eq.~(\ref{main_equations_continuous}) over $A'$, we get
\begin{eqnarray}\label{single_tone_Josephson_current}
2\cos\left(\phi'\right)\sin(\phi''/2) \approx 
\nonumber \\
\approx - 2\cos\left(A\theta'\sin\xi\right)\sin\left[\frac{A(\theta')^2}{2}\cos\xi\right] 
\nonumber \\
 +\cos(A\theta'\sin\xi)\cos\left[\frac{A(\theta')^2}{2}\cos\xi\right] 2A'\theta'\sin\xi + 
\nonumber \\
 + 2A'\cos(\xi)\sin(A\theta'\sin\xi)\sin\left[\frac{A(\theta')^2}{2}\cos\xi\right] \ .
\end{eqnarray}
There are two important points in further derivation. First, we use Jacobi-Anger identities~\cite{Abramovitz} and restrict ourselves to the contributions oscillating at the frequency of the drive $\omega$ neglecting, thus, the effects of higher harmonics. Second, for a rather strong amplitude of the drive $A$, the product $\tilde{A} = A\theta'$ can be of order unity and should be treated exactly, while the corrections to the leading-order equations are at least $\propto (\theta')^2 \ll 1$ and can be neglected. The above-mentioned arguments result in the following set of replacement rules applied to the right-hand-side of Eq.~(\ref{single_tone_Josephson_current}): 
\begin{subequations}\label{single_tone_replacement_rules}
\begin{align}
2\cos(A\theta'\sin\xi)\sin\left[\frac{A(\theta')^2}{2}\cos\xi\right] \to 
\nonumber ~~~~~~~~~~~~~~~~~~\\
 \to  2\cos(\xi) J_1(\tilde{A})A(\theta')^2/\tilde{A} \ ,~~~~~~~~~~~~~~~~~~~~~~~~~~~~~~~\\
2A'\theta' \sin(\xi)\cos\left(A\theta'\sin\xi\right)\cos\left[\frac{A(\theta')^2}{2}\cos\xi\right] \to 
\nonumber \\
 \to  2A'\theta'\sin(\xi)\left[J_0(\tilde{A}) -  J_2(\tilde{A})\right] \ ,~~~~~~~~~~~~~~~~\\
2A'\cos(\xi)\sin(A\theta'\sin\xi)\sin\left[\frac{A(\theta')^2}{2}\cos\xi\right] \to 
\nonumber ~~~~~~~~~~~~~\\
\to   2A'\theta'J_2(\tilde{A})\sin(\xi) ~~~~~~~~~~~~~~~~~~~~~~~~~~~~~~~\ ,
\end{align}
\end{subequations}
where $J_n(x)$ is the Bessel function of the first kind of order $n$~\cite{Abramovitz}.  
Combining Eqs.~(\ref{main_equations_continuous})-(\ref{single_tone_replacement_rules}), we obtain 
the reduced equations governing the spatial profiles of the amplitude $A(x)$ and phase $\theta(x)$ along the JTL
\begin{subequations}\label{single_tone_resulting_equations}
\begin{align}
\label{self_phase_modulation_equation}
(\theta')^2 = \frac{\omega^2/\omega_0^2}{\{1 + \alpha [2J_1(\tilde{A})/\tilde{A}-1]\}} \ ,\\
\label{single_tone_amplitude_equation}
A' = -  \frac{\gamma_{\omega}A\theta'}{\{1 + \alpha[ J_0(\tilde{A})-1]\}} \ .
\end{align}
\end{subequations}
Here, $\alpha = L/L_J=\beta_L/(1+\beta_L)$ is the inductance ratio, $\tilde A=A\theta'$, and we also introduce the dimensionless frequency-dependent dissipation parameter $\gamma_{\omega}=\omega L/2R_J$. The reduced equations (\ref{single_tone_resulting_equations}) are valid in the regime of $\gamma_\omega \ll 1$ and represent the main analytical result of this subsection.

\begin{figure}[tb]
\includegraphics[scale = 0.87]{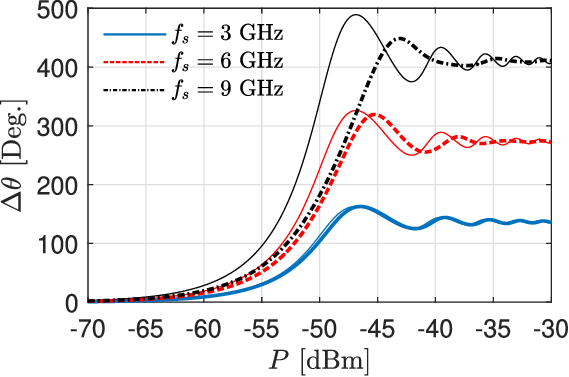}
\caption{Typical dependence of the phase length variation $\Delta\theta(P)-\Delta\theta(P\to 0)$ defined by Eq.~(\ref{phase_delay_definition}) on the microwave power $P$ for a  single applied microwave tone. The plots shown by thick lines are obtained using numerical solution of Eq.~(\ref{single_tone_resulting_equations}) in the presence of dissipation, while thin lines demonstrate the results for the lossless case ($\gamma_{\omega}=0$). The results were obtained for $\omega_0=649$~GHz and $\alpha= 0.34$. The dissipation parameter $\gamma_{\omega} = 0.004$, 0.008, and 0.012 for $f = 3$, 6, and 9~GHz, respectively.}
\label{Fig:phase_delay_comparison}
\end{figure}

Equation~(\ref{self_phase_modulation_equation}) describes the effect of the self-phase modulation and provides the implicit ($\tilde{A} = A\theta'$) local relation between the derivative of the phase and the amplitude of the response. Note that for arbitrary amplitude profile $A(x)$ in the transmission line, Eq.~(\ref{self_phase_modulation_equation}) can be solved iteratively for $\theta'$ by starting with the initial guess corresponding to $\theta' = \omega/\omega_0$, and  using this solution the \textit{phase length} of the propagating signal
\begin{equation}\label{phase_delay_definition}
\Delta\theta = \left[\theta(x = \ell)-\theta(x = 0)\right] 
\end{equation}
was calculated. One can clearly see that for small driving amplitudes (in the linear regime) $\Delta \theta \approx N\omega/\omega_0 = N\omega\sqrt{C_0 L}$, whereas in the limit $A\gg 1$ the ratio $J_1(\tilde{A})/\tilde{A}$ vanishes, and $\Delta \theta \approx [N\omega/\omega_0](1 - \alpha)^{-1/2} = N\omega \sqrt{C_0L_g}$, and, therefore, increasing the input microwave power of the drive should lead to a large increase in the phase length for long JTLs ($N \gg 1$). This effect can be attributed to a substantial variation of the total inductance of the unit cell from $L_gL_J/(L_g + L_J)$ in the linear regime to $L_g$ in the strongly nonlinear regime. Typical behavior of the phase length variation $\Delta\theta(P)-\Delta\theta(P\to 0)$ as a function of the power $P$ of an applied microwave radiation with frequencies $f = \omega/2\pi = 3$, 6, and 9~GHz and in the absence of dissipation ($\gamma_{\omega} = 0$), is presented in Fig.~\ref{Fig:phase_delay_comparison} by thin solid lines. Hereafter, the input power $P$ is taken to be the time-averaged power on $R_{\rm in} = 50$~Ohm resistor at the beginning of a JTL, which is related to the node phase amplitude at the starting node $A_0 = A(x = 0)$. Details of the derivation of the boundary condition are provided in Appendix~\ref{boundary_condition_derivation}. Note that in the absence of dissipation the node phase amplitude is a constant, which follows from Eq.~(\ref{single_tone_amplitude_equation}). The plots in Fig.~\ref{Fig:phase_delay_comparison} clearly show the pronounced oscillations in the phase length for large applied microwave power. The overall phase modulation increases with increasing frequency, which is in agreement with Eq.~(\ref{self_phase_modulation_equation}).

\begin{figure}[htpb]
\centering
\includegraphics[scale = 0.86]{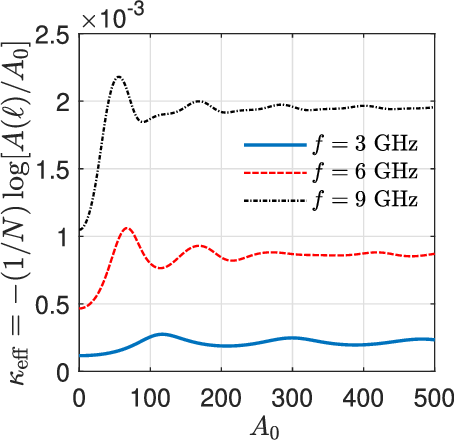}
\caption{Typical dependence of the effective decay constant $\kappa_{\rm eff}$ for electromagnetic waves on the input amplitude $A_0 = A(x = 0)$ for the single-tone case. The plots are obtained using numerical solution of Eqs.~(\ref{single_tone_resulting_equations}) for several frequencies of the drive $f = \omega/(2\pi)=3$, $6$, and $9$~GHz. The parameters are $\omega_0=649$~GHz, $\alpha=0.34$ and $\gamma_{\omega} = 0.004$, 0.008, and $0.012$ for $f = 3$, 6, and 9~GHz, respectively.}
\label{Fig:output_input_amplitude}
\end{figure} 

In the presence of a weak dissipation both the amplitude $A(x)$ and the local wave number $\theta'(x)$ vary along the JTL, and, therefore,
the phase length $\Delta \theta$ being an integral characteristic, depends on the spatial profile of the amplitude $A(x)$. The propagation of electromagnetic waves in weakly dissipative and strongly nonlinear JTLs can be quantitatively analyzed by making use of the large argument expansion of the Bessel functions in Eqs.~(\ref{single_tone_resulting_equations})  
\begin{subequations}\label{single_tone_wave_number_asymptotic_equation}
\begin{align}
    \label{single_tone_eikonal_asymptotics}
    (\theta')^2 = \frac{\omega^2}{\omega_0^2}\frac{1}{(1-\alpha)} \\
    \nonumber
     - \frac{\omega^2}{\omega_0^2}\frac{\alpha \pi}{(1-\alpha)^2}\left(\frac{2}{\pi\tilde{A}}\right)^{3/2}\cos(\tilde{A}-3\pi/4) \ ,\\
    \label{single_tone_amplitude_asymptotic_equation}
    A' = -\frac{\gamma_{\omega}\tilde{A}}{(1-\alpha)}\left[1 - \frac{\alpha}{(1-\alpha)}\frac{\cos(\tilde{A}-\pi/4)}{\sqrt{\pi\tilde{A}/2}}\right] \ .
\end{align}
\end{subequations}
Important features of the interplay between dissipation and strong nonlinearity can be understood from 
Eqs.~(\ref{single_tone_amplitude_equation}) and (\ref{single_tone_amplitude_asymptotic_equation}). Firstly, the signal amplitude decays inside the JTL and for the analysis of its behavior it is convenient to introduce the inverse effective decay length defined as $\kappa_{\rm eff} = -(1/N)\log[A(x = \ell)/A_0]$, where the input amplitude $A_0 \equiv A(x = 0)$. The Eqs~(\ref{single_tone_amplitude_equation}) and (\ref{single_tone_amplitude_asymptotic_equation}) imply the emergence of a spatially dependent crossover between the strongly nonlinear ($A(x)\gg 1$) and linear  ($A(x) \ll 1$)  regimes.  Secondly, one can expect that the inverse decay length $\kappa_{\rm eff}$ should be strongly enhanced from $\kappa_{\rm eff} \approx \gamma_\omega \omega/\omega_0$ ($A \ll 1$, the linear regime) up to $\kappa_{\rm eff} \approx \gamma_\omega (\omega/\omega_0)(1-\alpha)^{-3/2}$ ($A \gg 1$, the strongly nonlinear regime). This feature is demonstrated in Fig.~\ref{Fig:output_input_amplitude}, where we plot the inverse decay length $\kappa_{\rm eff}$ as a function of  the input amplitude $A_0$ for several signal frequencies $f_s = \omega_s/(2\pi) = 3$, 6.1, and 9~GHz. Note that in the strongly nonlinear regime numerical solution of Eqs.~(\ref{single_tone_resulting_equations}) gives somewhat smaller effective inverse decay length than the one predicted by the above estimate. Note also that the output amplitude also exhibits rather small oscillations as a function of the input one. In the limit $A\gg 1$ these oscillations are captured by Eq.~(\ref{single_tone_amplitude_asymptotic_equation}). 

To conclude this section we calculate the variation of the phase length in the presence of dissipation-induced propagation losses. Corresponding results obtained from numerical solutions of Eqs.~(\ref{single_tone_resulting_equations}) together with Eq.~(\ref{phase_delay_definition}) are presented in Fig.~\ref{Fig:phase_delay_comparison} by thick lines. One can clearly see that the propagation losses lead to smearing out phase length oscillations. This smearing becomes more pronounced at larger frequencies.

\subsection{\label{subsection:Sec_3_Two_Tone} Two-tone spectroscopy}

When two microwave tones are applied, i.e., both $P_s$ and $P_p$ are nonzero, we look for the solution of the nonlinear wave equation~(\ref{main_equations_continuous}) in the following form:
\begin{eqnarray}\label{two_tones_ansatz}
\phi(x,t) = \phi_s(x,t) + \phi_p(x,t) = 
\nonumber \\
=A_p(x)\cos[\omega_p t - \theta_p(x)] + A_s(x)\cos[\omega_s t - \theta_s(x)] \ ,
\end{eqnarray}
where $A_{p(s)}$, $\theta_{p(s)}$, $\omega_{p(s)}$ are the amplitude, phase and the frequency of the pump (signal) microwave tones. Here we consider the case of a relatively small amplitude of the microwave signal tone $A_s\ll A_p$, for which one can  neglect the back-action of a weak signal on a strong pump tone.

\begin{figure}[htpb]
\centering
\includegraphics[scale = 0.85]{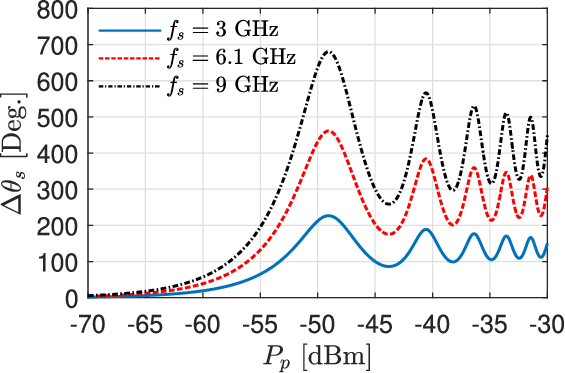}
\caption{Typical plots of the phase length variation $\Delta \theta_s(P_p) - \Delta \theta_s(P_p\to 0)$ for a weak probe signal as a function of the pump power $P_p$ for several frequencies of the signal $f_s = 3$, 6.1, and 9~GHz. Corresponding results are obtained using the solution of Eqs.~(\ref{two_tones_undepleated}) (lossless case). We choose the pump tone with the frequency $f_p = 6$~GHz and power $P_s = -100$ dBm. The parameters $\omega_0= 649$~GHz, $\alpha=0.34$, and $\gamma_s = \gamma_p = 0$.}
\label{Fig:two_tones_undepleated}
\end{figure}

We start with the analysis of the nonlinear wave equation (\ref{main_equations_continuous}) in the absence of losses (both dissipation parameters $\gamma_s = \omega_sL/(2R_J)$ and $\gamma_p = \omega_pL/(2R_J)$ are taken to be zero). The simplest approximation for the two-tone problem is to disregard the amplitude variations for both the pump and the signal. Such an approach provides a qualitative understanding of the behavior of the phase length for the weak signal. Substituting Eq.~(\ref{two_tones_ansatz}) with homogeneous $A_p$ and $A_s$ into Eq.~(\ref{main_equations_continuous}) and 
carrying out the derivation along the lines mentioned in the previous subsection, we derive the following set of equations:
\begin{subequations}\label{two_tones_undepleated}
\begin{align}
(\theta_p')^2 = \frac{\omega_p^2/\omega_0^2}{1 - \alpha[1-2J_1(\tilde{A}_p)/\tilde{A}_p]} \ ,\\
\label{eikonal_signal}
(\theta_s')^2 = \frac{\omega_s^2/\omega_0^2}{1 - \alpha[1-2J_0(\tilde{A}_p)J_1(\tilde{A}_s)/\tilde{A}_s]} \ ,
\end{align}
\end{subequations}
from which one can obtain the phase length for a probe signal by means of Eq.~(\ref{phase_delay_definition}). Here $\tilde{A}_s = A_s\theta_s'$ and $\tilde{A}_p = A_p\theta_p'$. The resulting Eq.~(\ref{eikonal_signal}) describes both the self- and cross-phase modulation for the signal propagating in a strongly driven medium. 
Typical dependencies of the cross-phase modulation on the pump power $P_p$ are shown in Fig.~\ref{Fig:two_tones_undepleated}. In the limit $A_s\ll 1$, the cross-phase modulation dominates. One can see that the resulting curves exhibit a large number of pronounced oscillations in the strongly nonlinear regime. The amplitude of such oscillations decreases with the power $P_p$ and increases with the frequency of the signal tone $\omega_s$. To illustrate the behavior of the phase length in the strongly nonlinear regime, one can make use of the large argument asymptotic of the Bessel function $J_0(\tilde{A}_p)$ in Eq.~(\ref{eikonal_signal}) and rewrite it in the form
\begin{eqnarray}\label{eikonal_signal_asymptotics}
(\theta_s')^2 = \frac{\omega_s^2}{\omega_0^2}\frac{1}{(1-\alpha)} \\
\nonumber
 - \frac{\omega_s^2}{\omega_0^2} \frac{\alpha}{(1-\alpha)^2}\frac{2J_1(\tilde{A}_s)}{\tilde{A}_s\sqrt{\pi\tilde{A}_p/2}}\cos\left(\tilde{A}_p - \pi/4\right) \ .
\end{eqnarray}
Comparison of Eqs.~(\ref{eikonal_signal_asymptotics}) and (\ref{single_tone_eikonal_asymptotics}) for the pump wave yields that the amplitude of the phase length oscillations associated with the self-phase modulation in the single-tone regime exhibits faster decay with increasing the pump power than the one for phase length oscillations of a weak signal in the two-tone regime induced by the cross-phase modulation.

\begin{figure}[htpb]
\centering
\includegraphics[scale = 0.87]{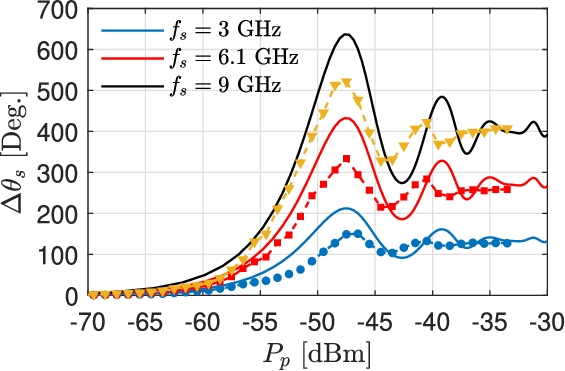}
\caption{Typical dependence of the cross-phase modulation, $\Delta \theta_s (P_p)-\Delta \theta_s (P_p \to 0)$ on the pump tone power $P_p$. The results shown by solid lines are obtained using numerical solutions of Eqs.~(\ref{weak_signal_system}) for the input power of the signal $P_s = -100$ dBm and the pump frequency $f_p = 6$ GHz. We chose the following parameters: $\omega_0= 649$~GHz, $\alpha=0.34$, $\gamma_p = 0.008$, $\gamma_s = 0.004$, 0.0082, and $0.012$ for $f_s = 3$, 6.1, 9~GHz, respectively. The symbols correspond to experimental results.}
\label{Fig:two_tones_with_dissipation_phase_delay}
\end{figure}

As a next step, we take the propagation losses into account. Substituting Eq.~(\ref{two_tones_ansatz}) into Eq.~(\ref{main_equations_continuous}), and considering the limit of a small amplitude of the signal microwave tone, we derive the following equation for the signal (see Appendix~\ref{modified_equations_two_tones_appendix} for details of the derivation):
\begin{equation}\label{weak_signal_equation}
\frac{1}{\omega_0^2}\ddot{\phi}_s - q_p(x)\frac{\partial^2\phi_s}{\partial x^2}-\frac{L}{R_J}\frac{\partial^2\dot{\phi}_s}{\partial x^2}+\Gamma_p(x)\frac{\partial \phi_s}{\partial x} = 0 \ ,
\end{equation}
where we introduced 
\begin{subequations}
\begin{align}
q_p(x) = 1 - \alpha\left[1 - J_0(\tilde{A}_{p})\right] \ ,\\
\Gamma_p(x) = \alpha A_p'\theta_p'J_1(\tilde{A}_{p}) \ ,
\end{align}
\end{subequations}
and the spatial profile of the pump wave in JTL is obtained via the solution of Eqs.~(\ref{single_tone_resulting_equations}). One can see that Eq.~(\ref{weak_signal_equation}) reveals the additional dissipation-induced coupling of the signal wave to the pump described by the function $\Gamma_p(x)\propto A_p'$. Despite the fact that for a weak dissipation the derivative $A_p'$ is rather small, we find that the above-mentioned coupling cannot be neglected within the parameter range of interest. Substituting the solution of the form $\phi_s(x,t) = A_s(x)\cos(\omega_s t - \theta_s(x))$ into Eq.~(\ref{weak_signal_equation}), we derive the following system of equations:
\begin{subequations}\label{weak_signal_system}
\begin{align}
-\left(\frac{\omega_s}{\omega_0}\right)^2 A_s  - q_p\left[A_s'' - A_s(\theta_s')^2\right] \\
\nonumber
 - 2\gamma_s(A_s\theta_s'' + 2A_s'\theta_s') + \Gamma_pA_s' = 0 \ ,\\
-q_p\left[A_s\theta_s'' + 2A_s'\theta_s'\right] + 2\gamma_s\left[A_s'' - A_s(\theta_s')^2\right] \\
\nonumber
 + \Gamma_pA_s\theta_s' = 0 \ ,
\end{align}
\end{subequations}
which we solve to determine the phase length in the presence of dissipation. For this purpose, we first solve Eqs.~(\ref{single_tone_resulting_equations}) for the pump wave and obtain the functions $q_p(x)$ and $\Gamma_p(x)$. Second, we substitute these functions into (\ref{weak_signal_system}) 
and perform numerical integration. We find the evanescent solutions for the signal wave using the following boundary conditions (see Appendix~\ref{modified_equations_two_tones_appendix} for the corresponding discussion):
\begin{subequations}\label{weak_signal_boundary_conditions}
\begin{align}
A_s'(0) = {\rm Re}\left(\lambda_-\right)A_s(0) \ , \ \ \theta_s'(0) = -{\rm Im}\left(\lambda_-\right) \ ,\\
\lambda_{-} = \frac{\Gamma_p(0)}{2[q_p(0) +2 i\gamma_s]} \\
\nonumber
 - i\sqrt{\frac{(\omega_s/\omega_0)^2}{[q_p(0)+2i\gamma_s]}-\frac{[\Gamma_p(0)]^2}{4[q_p(0)+2i\gamma_s]^2}} \ ,
\end{align}
\end{subequations}
where $q_p(0) = q_p(x = 0)$ and $\Gamma_p(0) = \Gamma_p(x=0)$. As a result, we obtain the dependencies of the phase length for a weak probe signal on the input power of the pump $P_p$. The resulting plots shown in Fig.~\ref{Fig:two_tones_with_dissipation_phase_delay} demonstrate that the dissipation-induced propagation losses are responsible for smearing out oscillations of the phase length in the two-tone case.

It is interesting to note that the resulting $\Delta\theta_s(P_p)$ plots can be well approximated 
by the solution of the reduced equations for the signal. Indeed, neglecting the second derivatives in Eqs.~(\ref{weak_signal_system}), we get the following expression:
\begin{equation}
\theta_s' = \frac{\Gamma_p\gamma_s}{2q_p^2} \pm \sqrt{\frac{1}{q_p}\left(\frac{\omega_s}{\omega_0}\right)^2 + \left[\frac{\Gamma_p\gamma_s}{2q_p^2}\right]^2 - \frac{\Gamma_p^2}{2q_p^2}} \ ,
\end{equation}
the integration of which gives us $\Delta\theta_s(P_p)$ deviating from the exact one by several percent.

Finally, let us briefly discuss the comparison between our experimental and theoretical results (see Fig.~\ref{Fig:two_tones_with_dissipation_phase_delay}). Note that the dissipation parameters used in Fig.~\ref{Fig:two_tones_with_dissipation_phase_delay} represent the best fit to experimental results. One can see that overall agreement between theory and experiment is good and the proposed theoretical description captures well experimentally observed behavior for small and large pump powers as well as the number of pronounced oscillations of the phase length variation. Still, it can be seen that theoretical description overestimates the oscillation amplitude of the $\Delta\theta_s(P_p)$ curves. This discrepancy can be attributed to the processes leading to a substantial suppression of the pump amplitude deep inside a JTL, e.g., the harmonic generation and parametric amplification of the signal, which were not taken into account in our model. Our typical experimental results on $|S_{21}|$ as a function of the pump power demonstrate a weak parametric amplification in the strongly nonlinear regime (it is presented in Appendix~\ref{experimental_gain}). 

\section{\label{section:Sec_6_discussion}Discussion and conclusion} 

In conclusion, we present experimental and theoretical studies of microwave propagation in Josephson transmission lines (JTLs). A JTL consists of an array of rf-SQUIDs galvanically coupled to a low-dissipative coplanar waveguide. Performing two-tone spectroscopy, we measured the phase-length variation of the microwave signal as a function of the input power of the pump tone. We were able to experimentally achieve the strongly nonlinear regime in which the phase length is strongly enhanced and exhibits a few pronounced oscillations as the pump power increases. The phase length saturates in the limit of a high input pump power, and the cross-phase modulation increases with increasing probe frequency (see Fig.~\ref{fig:TT}).

Experimental observations are quantitatively explained in the framework of the electrodynamic model, in which low-dissipative rf-SQUIDs  are directly embedded into the dissipationless transmission line (see schematic in Fig.~\ref{fig:circuit1}). By making use of the elaborated electromagnetic model we derive the nonlinear wave equation governing the electromagnetic wave propagation in such JTLs (Eq. (\ref{main_equations_continuous})).  The main idea borrowed from Refs. \cite{Multistability,Kiselev} is that for a strong driving tone the response of the JTL is mostly dominated by the signal at the frequency of the drive. This idea allowed us to solve  Eq.~(\ref{main_equations_continuous})  approximately and to obtain the dependence of the phase length, $\Delta \theta(P_s,P_p)$ on the input powers of the pump and signal tones (see Figs.~\ref{Fig:phase_delay_comparison} and~\ref{Fig:two_tones_with_dissipation_phase_delay}).

It is shown that the experimentally observed pronounced oscillations of the phase length reflect the oscillatory dependence of the Josephson inductances as a function of the pump power. We demonstrate that the number of oscillations observed in the dependence of the phase length vs. pump power is highly sensitive to dissipation-induced propagation losses in the JTL. Taking into account the unavoidable dissipation, we uncover the peculiar interplay between the dissipation and strong nonlinearity in JTLs. It has been demonstrated that (i) the propagation losses of the pump wave can be enhanced in the strongly nonlinear regime, (ii) the oscillations of the phase length are smeared out even in the presence of a weak dissipation (compare Figs.~\ref{Fig:two_tones_undepleated} and \ref{Fig:two_tones_with_dissipation_phase_delay}). The latter feature originates from the decay of the pump amplitude $A_p(x)$ deep inside the JTL as $\ell \gg (1/\kappa_{\rm eff}) \simeq [2R_J\omega_0/(\omega^2 L)] (1-\alpha)^{3/2}$. Therefore, the oscillations should be recovered for short JTLs as $\ell \leq (1/\kappa_{\rm eff})$.

Finally, it is important to note that there are several other mechanisms, such as higher-harmonic generation and parametric amplification, which can affect the experimentally observed behavior of the phase length as a function of pump power. Since the phase length in both the single- and two-tone cases is an integral characteristic, which in the nonlinear regime depends on the spatial profile of the wave amplitude, any mechanism leading to appreciable pump depletion can produce similar effects on its behavior. In particular, our analysis presented in Appendix~\ref{app:third_harmonic} shows that, in the strongly nonlinear regime, higher-harmonic generation need not remain negligible and the interaction between the pump tone and its harmonics can become appreciable. Nevertheless, the reduced one-harmonic approximation for the pump still captures the leading behavior of the phase length rather well. By contrast, a quantitative description of the output wave amplitudes is expected to require a multimode theory, since dissipation-induced propagation losses, higher-harmonic generation, and intermodulation products can all significantly modify pump depletion and wave mixing. These processes can, therefore, provide important contributions to gain compression in JTWPAs. A quantitative treatment of harmonic generation and wave mixing in JTLs lies beyond the scope of the present work and represents an important direction for future research. The studies presented in this work are an important step toward understanding the response of strongly driven JTLs, which are promising for the design of travelling-wave parametric amplifiers and other superconducting devices exploiting strongly nonlinear phenomena.

Raw experimental data on the transmission coefficient in the two-tone regime are publicly available~\cite{Zenodo}.

\section{Acknowledgments}
The authors acknowledge Alexey Ustinov for helpful discussions and comments on the
manuscript. The work was supported by the Ministry of Science and Higher Education of Russian Federation within the program Priority-2030 (Strategic Technology Project of NUST MISIS "Quantum Internet") and by the Ministry of Science and Higher Education of Russian Federation (MIPT Shared Facilities Center).

\appendix
\begin{widetext}

\section{Derivation of the boundary condition}\label{boundary_condition_derivation}
In this section we provide the derivation of the boundary condition, which relates the time-averaged input power to the amplitude of the node phase variable at the starting node of the circuit. For this purpose we write down the equation expressing the current conservation condition at the starting node of the circuit (equivalent circuit is shown in Fig.~\ref{Fig:starting_node_equivalent_circuit})
\begin{eqnarray}\label{current_conservation_starting_node}
C_0\ddot{\phi}_1 + \frac{\dot{\phi}_1}{R_i}-C_J(\ddot{\phi}_2 - \ddot{\phi}_1)-\frac{(\phi_2 - \phi_1)}{L_g} -\frac{1}{L_J}\sin(\phi_2- \phi_1)-\frac{(\dot{\phi}_2 - \dot{\phi}_1)}{R_J} = I_{\rm in}(t) \ .
\end{eqnarray}
Here $I_{\rm in}(t)  = V_{\rm in}(t)2\pi/\Phi_0R_i$, $V_{\rm in}(t)$ is the driving voltage at the voltage source, and $R_i = 50$~Ohm is the load resistor. As in the main text, here we consider the low-frequency response and neglect the effects of the junction capacitance~$C_J$. 

\begin{figure}[htpb]
\centering
\includegraphics[scale = 0.8]{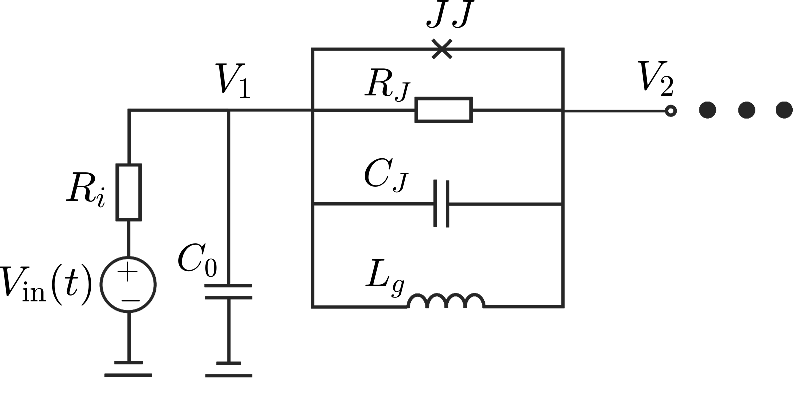}
\caption{Equivalent scheme of the circuit used to derive the relation between the time-averaged input power and the amplitude of the node phase variable at the starting node.}
\label{Fig:starting_node_equivalent_circuit}
\end{figure} 

Considering a single tone with $V_{\rm in}(t) = V_p\cos(\omega t)$, we look for the solution of the above equation in the form
\begin{equation}\label{starting_node_ansatz}
\phi_1 = A_1\cos(\omega t - \theta_1) \ , \ \ \phi_2 = A_2\cos(\omega t - \theta_2) \ ,
\end{equation}
where $V_1 = \dot{\phi}_1\Phi_0/(2\pi)$ and $V_2 = \dot{\phi}_2\Phi_0/(2\pi)$. Substituting Eq.~(\ref{starting_node_ansatz}) into Eq.~(\ref{current_conservation_starting_node}) and keeping the terms oscillating at the frequency of the drive
\begin{subequations}
\begin{align}
\phi_2 - \phi_1 \approx \phi' \approx A\theta'\sin(\xi) + A'\cos(\xi) \ ,\\
\sin(\phi_2 - \phi_1)\approx \sin\left[A\theta'\sin\xi + A'\cos\xi\right] \approx 
 \sin(A\theta'\sin\xi)+A'\cos\xi\cos(A\theta'\sin\xi) \to\\
\nonumber
\to 2J_1(A\theta')\sin(\xi) 
 + A'\cos(\xi)\left[J_0(A\theta')+J_2(A\theta')\right] \ ,
 \ \ 
\end{align}
\end{subequations}
we arrive at the following equation:
\begin{eqnarray}\label{amplitude_phase_shift_system}
\cos(\xi)\left\{-\frac{\omega^2}{\omega_0^2}A - A' + \alpha A'\left[1 - J_0(\tilde{A})-J_2(\tilde{A})\right]-2\gamma_{\omega}\tilde{A}\right\} \\
\nonumber
+\sin(\xi)\left\{-2\gamma_iA-\tilde{A} + \alpha \left[\tilde{A} - 2J_1(\tilde{A})\right]+2\gamma_{\omega}A'\right\} = A_{\rm dr}\cos(\omega t) \ .
\end{eqnarray}
Here $\xi = \omega t - \theta$, $\theta = \theta(x = 0)$, $A = A(x = 0)$, $A' = A'(x = 0)$, $\tilde{A} = A\theta'$,  $\gamma_i = \omega L/2R_i$, $\gamma_{\omega} = \omega L/2R_J$, and $A_{\rm dr} = V_p 2\pi L/\Phi_0 R_i$. From Eq.~(\ref{amplitude_phase_shift_system}) we get the following system, from which one can determine the signal amplitude and the phase shift of the steady-state solution at the starting node of the circuit   
\begin{subequations}\label{amplitude_phase_shift_solution}
\begin{align}
\tan(\theta) = \frac{-\left\{2\gamma_iA + \tilde{A} - \alpha[\tilde{A}-2J_1(\tilde{A})]-2\gamma_{\omega}A'\right\}}{\left\{\frac{\omega^2}{\omega_0^2}A + A'\left[1 - \alpha(1-J_0(\tilde{A})-J_2(\tilde{A}))\right] +2\gamma_{\omega}\tilde{A}\right\}} \ ,\\
-\cos(\theta)\left\{\frac{\omega^2}{\omega_0^2}A + A'\left[1 - \alpha(1-J_0(\tilde{A})-J_2(\tilde{A}))\right] + 2\gamma_{\omega}\tilde{A}\right\} \\
\nonumber
+ \sin(\theta)\left\{2\gamma_iA + \tilde{A} - \alpha[\tilde{A} - 2J_1(\tilde{A})]-2\gamma_{\omega}A'\right\} = A_{\rm dr} \ .
\end{align}
\end{subequations}
Solving the above system, we obtain the following relation between the amplitude of the drive $A_{\rm dr}$ and amplitude of the node flux variable at the starting node of the circuit 
\begin{equation}\label{A_Adr_relation}
A_{\rm dr} = \sqrt{\left\{A\left(\frac{\omega}{\omega_0}\right)^2 + A' + 2\gamma_{\omega}\tilde{A}-\alpha A'\left[1-J_0(\tilde{A})-J_2(\tilde{A})\right]\right\}^2 + \left\{\tilde{A} + 2\gamma_i A - 2\gamma_{\omega} A' - \alpha\left[\tilde{A}-2J_1(\tilde{A})\right]\right\}^2} \ ,
\end{equation}
where $\theta'$ and $A'$ at the starting node should be obtained from Eqs.~(\ref{single_tone_resulting_equations}).

As a next step, we use Eqs.~(\ref{amplitude_phase_shift_solution}) and (\ref{A_Adr_relation}) to calculate the relation between the input power averaged over the radiation period ($T = 2\pi/\omega$) and the amplitudes $A$ and $A_{\rm dr}$. The time-averaged power at the load resistor reads as
\begin{equation}
P = \frac{1}{T}\left(\frac{\Phi_0}{2\pi}\right)^2\int_0^{T}dt \frac{\dot{\phi}_1\left[\dot{\phi}_1 - 2\pi V_{\rm in}(t)/\Phi_0\right]}{R_i} = \left(\frac{\Phi_0}{2\pi}\right)^2\frac{1}{2R_i}\left[\omega^2A^2 - \omega A A_{\rm dr}(A)\frac{R_i}{L}\sin(\theta)\right] \ ,
\end{equation}
where we explicitly indicated the dependence of the drive amplitude on the node phase amplitude at the starting node $A_{\rm dr}(A)$. Within the parameter range considered in our work, the dissipation is small, so we neglect the terms $\propto \gamma_{\omega}$ and the derivatives $A'$ in Eqs.~(\ref{amplitude_phase_shift_solution}) and (\ref{A_Adr_relation}). For typical parameters we find that $\theta \approx - \pi/2$, so the relation between the input power and the node phase amplitude is given by the following expression:
\begin{equation}\label{P_A_expression}
P = \left(\frac{\Phi_0}{2\pi}\right)^2\frac{1}{2R_i}\left[\omega^2A^2 + \omega A A_{\rm dr}(A)\frac{R_i}{L}\right] \ .
\end{equation}
Finally, we substitute the considered range of boundary conditions $A = A(x = 0)$ used in our calculations into Eq.~(\ref{A_Adr_relation}) to find $A_{\rm dr}(A)$ dependence. Next, both $A$ and corresponding $A_{\rm dr}(A)$ function are then substituted into Eq.~(\ref{P_A_expression}), which yields the time-averaged input power $P(A)$ corresponding to a given set of boundary conditions.

\section{Derivation of Eqs.~(\ref{weak_signal_equation}) and~(\ref{weak_signal_boundary_conditions}) in the main text}\label{modified_equations_two_tones_appendix}

In this section we provide the detailed derivation of the linearized equation for the weak signal in the two-tone case together with the corresponding boundary conditions~(\ref{weak_signal_boundary_conditions}).  
Substituting the sum $\phi = \phi_s + \phi_p$ into Eq.~(\ref{single_tone_Josephson_current}) and keeping the terms linear in the amplitude of the signal, we get
\begin{eqnarray}
 2\cos\left(\frac{\partial\phi}{\partial x}\right)\sin\left(\frac{1}{2}\frac{\partial^2\phi}{\partial x^2}\right) \approx\\
\nonumber
\approx 2\cos\left(\frac{\partial \phi_p}{\partial x}\right)\sin\left(\frac{1}{2}\frac{\partial^2\phi_p}{\partial x^2}\right) + \frac{\partial^2\phi_s}{\partial x^2}\cos\left(\frac{\partial\phi_p}{\partial x}\right)\cos\left(\frac{1}{2}\frac{\partial^2\phi_p}{\partial x^2}\right) - 2\frac{\partial \phi_s}{\partial x}\sin\left(\frac{\partial\phi_p}{\partial x}\right)\sin\left(\frac{1}{2}\frac{\partial^2\phi_p}{\partial x^2}\right) \ .
\end{eqnarray}
The first term in the right-hand-side of the above expression can only contribute to the equations for the pump. Regarding the equations for the signal, it is clear that the coupling functions to the pump wave should be given by time averages of the products $\cos(\phi_p')\cos(\phi_p''/2)$ and $\sin(\phi_p')\sin(\phi_p''/2)$. Using the adopted form of the node flux variable for the pump $\phi_p(x,t) = A_p(x)\cos[\omega_p t - \theta_p(x)]$, we get the following combinations
\begin{subequations}\label{combinations}
\begin{align}
\phi_p' - \phi_p''/2 \approx \left[A_p' + A_p(\theta_p')^2/2\right]\cos(\xi_p) + (A_p\theta_p'-A_p'\theta_p')\sin(\xi_p) \ ,\\
\phi_p' + \phi_p''/2 \approx \left[A_p'-A_p(\theta_p')^2/2\right]\cos(\xi_p)+(A_p\theta_p' + A_p'\theta_p')\sin(\xi_p) \ .
\end{align}
\end{subequations}
Here $\xi_p = \omega_p t - \theta_p$. It is convenient to introduce the angles
\begin{subequations}
\begin{align}
\cos(\eta_-) = (A_p\theta_p'-A_p'\theta_p')/\tilde{A}_{p-} \ , \ \ \sin(\eta_-) = \left[A_p' + A_p(\theta_p')^2/2\right]/\tilde{A}_{p-} \ ,\\
\tilde{A}_{p-} = \sqrt{\left[A_p'+A_p(\theta_p')^2/2\right]^2 + \left(A_p\theta_p' -A_p'\theta_p'\right)^2} \ ,\\
\cos(\eta_+) = (A_p\theta_p' + A_p'\theta_p')/\tilde{A}_{p+} \ , \ \ \sin(\eta_+) = \left[A_p'-A_p(\theta_p')^2/2\right]/\tilde{A}_{p+} \ ,\\
\tilde{A}_{p+} = \sqrt{\left[A_p'-A_p(\theta_p')^2/2\right]^2 + (A_p\theta_p'+A_p'\theta_p')^2} \ ,
\end{align}
\end{subequations}
and cast the combinations in Eqs.~(\ref{combinations}) to more compact form:
\begin{equation}
\phi_p'-\phi_p''/2 = \tilde{A}_{p-}\sin(\xi_p + \eta_{-}) \ , \ \ 
\phi_p'+\phi_p''/2 = \tilde{A}_{p+}\sin(\xi_p + \eta_+) \ .
\end{equation}
The time-averaged products can be obtained as follows:
\begin{subequations}
\begin{align}
\cos(\phi_p')\cos\left(\frac{1}{2}\phi_p''\right) = \frac{1}{2}\left\{\cos[\tilde{A}_{p-}\sin(\xi_p + \eta_-)] + \cos\left[\tilde{A}_{p+}\sin(\xi_p + \eta_+)\right]\right\} \to \frac{1}{2}\left[J_0(\tilde{A}_{p-}) + J_0(\tilde{A}_{p+})\right] \ ,\\
\sin(\phi_p')\sin\left(\frac{1}{2}\phi_p''\right) = \frac{1}{2}\left\{\cos[\tilde{A}_{p-}\sin(\xi_p + \eta_-)] - \cos\left[\tilde{A}_{p+}\sin(\xi_p + \eta_+)\right]\right\} \to \frac{1}{2}\left[J_0(\tilde{A}_{p-}) - J_0(\tilde{A}_{p+})\right] \ .
\end{align}
\end{subequations}
As a next step we expand the Bessel functions over the derivative $A_p'$. We get the following terms  
\begin{subequations}
\begin{align}
\tilde{A}_{p+} \approx \tilde{A}_{p} + A_pA_p'(\theta_p')^2/2\tilde{A}_{p} \ , \ \ \tilde{A}_{p-}\approx \tilde{A}_{p}-A_pA_p'(\theta_p')^2/2\tilde{A}_{p} \ ,\\
\tilde{A}_{p} = A_p\sqrt{(\theta_p')^2 + (\theta_p')^4/4} \approx A_p \theta_p' \ .
\end{align}
\end{subequations}
The leading-order terms in the expansion of the time averaged products are as follows:
\begin{subequations}
\begin{align}
\cos(\phi_p')\cos\left(\phi_p''/2\right) \to J_0(\tilde{A}_{p}) \ ,\\
\sin(\phi_p')\sin\left(\phi_p''/2\right) \to J_1(\tilde{A}_{p}) A_pA_p'(\theta_p')^2/2\tilde{A}_{p} \approx A_p'\theta_p' J_1(\tilde{A}_{p})/2 \ .
\end{align}
\end{subequations}
As a result, the dynamics of the weak signal in the presence of the pump wave is governed by the following equation: 
\begin{eqnarray}
C_0\frac{\partial^2\phi_s}{\partial t^2} -\frac{1}{L_g}\frac{\partial^2\phi_s}{\partial x^2}-\frac{1}{R_J}\frac{\partial^3\phi_s}{\partial t\partial x^2} = \frac{1}{L_J}\left[J_0(\tilde{A}_{p})\frac{\partial^2\phi_s}{\partial x^2} - A_p'\theta_p'J_1(\tilde{A}_{p})\frac{\partial\phi_s}{\partial x}\right] \ .
\end{eqnarray}
The above equation can be cast to the form presented by Eq.~(\ref{weak_signal_equation}) in the main text.
\end{widetext}

To derive the boundary conditions~(\ref{weak_signal_boundary_conditions}), we consider Eq.~(\ref{weak_signal_equation}) in the vicinity of the initial point of the circuit ($x = 0$) and take $\phi_s(x,t) = \phi_{s\omega}(x)e^{i\omega_s t}$. We get the following equation
\begin{eqnarray}
\left[q_p(0) + 2i\gamma_s\right]\frac{d^2\phi_{s\omega}}{dx^2} \\
\nonumber
-\Gamma_p(0)\frac{d\phi_{s\omega}}{dx} + \left(\frac{\omega_s}{\omega_0}\right)^2\phi_{s\omega} = 0 \ .
\end{eqnarray}
The solutions of the above equation can be presented in the form $\phi_{s\omega}(x) = \phi_{s\omega}(0)e^{\lambda x}$ with the exponents
\begin{eqnarray}
\lambda_{\pm} = \frac{\Gamma_p(0)}{2[q_p(0) + 2i\gamma_s]}\\
\nonumber
\pm i\sqrt{\frac{(\omega_s/\omega_0)^2}{[q_p(0)+2i\gamma_s]}-\frac{[\Gamma_p(0)]^2}{4[q_p(0)+2i\gamma_s]^2}} \ .
\end{eqnarray}

It is straightforward to show that the solution, which decays to the right $\phi_{s\omega}(x) = \phi_{s\omega}(0)e^{\lambda_- x}$. 
 Separating the real and imaginary parts $\lambda_- = \lambda_-' + i\lambda_{-}''$, we find
\begin{equation}
\phi_s(x,t) = A_s(0)e^{\lambda_-'x}\cos(\omega_s t + \lambda_-'' x) \ ,
\end{equation}
and obtain the necessary boundary conditions
\begin{equation}
A_s'(0) = \lambda_-'A_s(0) \ , \ \ \theta_s'(0) = -\lambda_-'' \ ,
\end{equation}
which can be cast to the form presented by Eqs.~(\ref{weak_signal_boundary_conditions}) in the main text.

\section{Typical experimental results for $|S_{21}|$}\label{experimental_gain}

\begin{figure}[htpb]
\centering
\includegraphics[scale = 0.96]{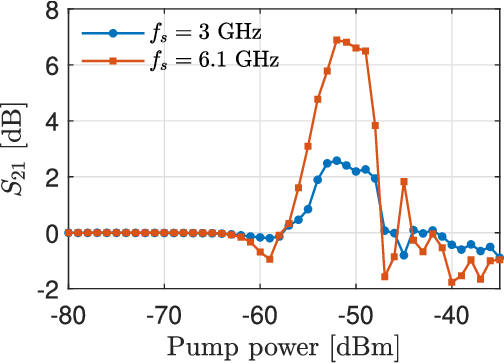}
\caption{Typical dependencies of $|S_{21}|$ for a weak probe signal as a function of the pump power obtained from experimental measurements. Data were obtained for the pump frequency $f_p = 6$~GHz and the signal frequencies $f_s = 3$ and 6.1~GHz.}
\label{Fig:experimental_gain}
\end{figure} 

In this section we briefly discuss experimental results for the transmission coefficient $S_{21}$~[dB] as a function of the pump power. Corresponding plots shown in Fig.~\ref{Fig:experimental_gain} were obtained for the pump frequency $f_p = 6$~GHz and several frequencies of the signal $f_s = 3$ and 6.1~GHz. These results demonstrate the main features of amplification in the strongly nonlinear regime. One can clearly see that in the strongly nonlinear regime the gain is rather moderate. For example, the results in Fig.~\ref{Fig:experimental_gain} reveal the maximum gain for $f_s = 3$ and 6.1~GHz of approximately 2.5 and 7 dB, respectively. For sufficiently large pump powers $S_{21}$~[dB] changes sign and exhibits oscillatory behavior.

\section{Third-harmonic generation and limitations of the one-harmonic approximation }
\label{app:third_harmonic}

In this section we consider third-harmonic generation within a simplified extension of the theoretical approach developed in the main text. This analysis is aimed both at estimating the magnitude of the third harmonic generated by a strong principal tone and at clarifying the limitations of the reduced one-harmonic approximation in the strongly nonlinear regime. We write the total phase field as
\begin{equation}
\phi(x,t)=\phi_1(x,t)+\phi_3(x,t) \ ,
\end{equation}
where $\phi_1 = A_1\cos(\omega t - \theta_1)$ is the strong principal tone at frequency $\omega$, while $\phi_3$ is the third harmonic generated by the nonlinearity. In the following we represent the third-harmonic contribution in the form
\begin{equation}
\phi_3(x,t)=\frac{1}{2}\left[u_3(x)e^{-3i\omega t}+u_3^*(x)e^{3i\omega t}\right].
\label{eq:D2}
\end{equation}
As a next step, we substitute Eq.~(\ref{eq:D2}) into Eq.~(\ref{main_equations_continuous}), linearize the nonlinear wave equation with respect to $\phi_3$ and neglect the back action of the third harmonic on the principal tone. We get as a result
\begin{eqnarray}
\left(q_0-6i\gamma_\omega\right)u_3''-\Gamma_0 u_3'
+(9\omega^2/\omega_0)^2u_3 \ \ \ \\
\nonumber
+\alpha J_6(\tilde A_1)\left(u_3''\right)^*e^{6i\theta_1}
+6i\alpha K_1 J_6(\tilde A_1)\left(u_3'\right)^*e^{6i\theta_1}
=\\
\nonumber
=6\alpha K_1 J_3(\tilde A_1)e^{3i\theta_1} \ .
\label{eq:D3}
\end{eqnarray}
Here
\begin{equation}
q_0 = 1-\alpha+\alpha J_0(\tilde A_1) \ , \ \ 
K_1=\theta_1' \ ,
\label{eq:D4}
\end{equation}
$\tilde A_1 = A_1K_1$, $\theta_1'$ is given by the solution of Eq.~(\ref{self_phase_modulation_equation}) with $A = A_1$, $J_3(x)$ and $J_6(x)$ denote the Bessel functions of the first kind of order 3 and 6, respectively. It is important to note that the simplified treatment presented below is valid in the regime $|u_3|/A_1\ll 1$. In the opposite case, quantitative analysis requires taking account of the back action of the third harmonic on the principal tone and the coupling to other harmonics.

\begin{figure}[htpb]
\centering
\includegraphics[scale = 0.96]{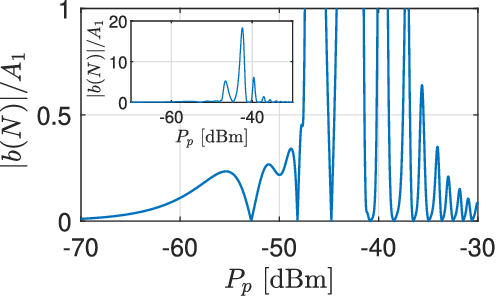}
\caption{Typical dependence of the ratio $|b(N)|/A_1$ as a function of the microwave power of the principal tone. Here $|b(N)|$ is given by Eq.~(\ref{eq:D11}) and represents a rough estimate of the third harmonic amplitude at the line output, while $A_1$ is the amplitude of the principal tone. The data were obtained for $\omega/2\pi = 6$~GHz, and the circuit parameters are similar to the ones used in the main text. The estimate shown in this figure is based on the assumption $|u_3|/A_1 \ll 1$. The regime $|b(N)|/A_1 \gtrsim 1$ lies beyond the validity of the underlying approximation.}
\label{Fig:third_harmonic_estimate}
\end{figure}

For a crude analytical estimate, we consider the lossless case and take a spatially homogeneous principal tone, so that $\gamma_\omega=0$, $\Gamma_0=0$, and $\theta_1(x)=K_1 x$ with constant $K_1$. Treating Eq.~(\ref{eq:D4}) within the slowly varying envelope approximation, we introduce 
\begin{equation}
u_3(x)=b(x)e^{3iK_1x} \ ,
\label{eq:D5}
\end{equation}
and obtain
\begin{equation}
6iQ\,b' + \delta\, b + \varkappa\, b^* = S_3,
\label{eq:D6}
\end{equation}
where $Q=q_0K_1$, $\delta = 9\left(\omega^2/\omega_0^2-q_0K_1^2\right)$, $\varkappa = 9\alpha K_1^2 J_6(\tilde A_1)$, and $S_3 = 6\alpha K_1 J_3(\tilde A_1)$.
Imposing the boundary condition
\begin{equation}
b(0)=0 \ ,
\label{eq:D8}
\end{equation}
which corresponds to the absence of an injected third harmonic at the input of the line and writing $b(x)=X(x)+iY(x)$,  Eq.~(\ref{eq:D6}) reduces to
\begin{equation}
X''+\Lambda^2 X = \Lambda^2 X_p \ ,
\label{eq:D9}
\end{equation}
with $\Lambda=\sqrt{\delta^2-\varkappa^2}/6Q$ and $X_p=S_3/(\delta+\varkappa)$.
Solving Eqs.~(\ref{eq:D9}) together with the boundary condition (\ref{eq:D8}), we find
\begin{equation}
b(x)=
X_p\bigl[1-\cos(\Lambda x)\bigr]
-iX_p\sqrt{\frac{\delta+\varkappa}{\delta-\varkappa}}\sin(\Lambda x).
\label{eq:D11}
\end{equation}
The corresponding finite-length estimate for the third-harmonic amplitude at the output of the transmission line within the simplified approach is $|b(N)|$. Typical behavior of the ratio $|b(N)|/A_1$ as a function of the microwave power of the principal tone is shown in Fig.~\ref{Fig:third_harmonic_estimate}. This plot clearly indicates that the initial assumption underlying the simplified description $|u_3|/A_1\ll 1$ breaks down in the power range approximately from $-50$ to $-35$~dBm and suggests that in the strongly nonlinear regime the third harmonic need not remain negligible. In this regime one can also expect an appreciable interaction between the principal tone and its higher harmonics. Nevertheless, the reduced one-harmonic description remains useful for capturing the leading behavior of the phase length, while a quantitative treatment of higher harmonics and their mutual coupling requires a separate study.

\bibliography{bibliography}

\end{document}